\documentclass[aps,prd,preprint,superscriptaddress,showpacs,preprintnumbers]{revtex4-1}
\usepackage[colorlinks=true, pdfstartview=FitV, linkcolor=red, citecolor=blue, urlcolor=black, pdftitle={},pdfauthor={Tomoya Hayata},pdfsubject={}, pdfkeywords={}]{hyperref}
\usepackage[pdftex]{graphicx}
\usepackage{setspace,bm}
\usepackage{latexsym,amssymb,amsmath,mathrsfs}
\usepackage{color}
\makeatletter
\providecommand \@ifxundefined [1]{%
 \@ifx{#1\undefined}
}%
\providecommand \@ifnum [1]{%
 \ifnum #1\expandafter \@firstoftwo
 \else \expandafter \@secondoftwo
 \fi
}%
\providecommand \@ifx [1]{%
 \ifx #1\expandafter \@firstoftwo
 \else \expandafter \@secondoftwo
 \fi
}%
\providecommand \bibnamefont  [1]{#1}%
\providecommand \bibfnamefont [1]{#1}%
\providecommand \citenamefont [1]{#1}%
\providecommand \href@noop [0]{\@secondoftwo}%
\providecommand \href [0]{\begingroup \@sanitize@url \@href}%
\providecommand \@href[1]{\@@startlink{#1}\@@href}%
\providecommand \@@href[1]{\endgroup#1\@@endlink}%
\providecommand \@sanitize@url [0]{\catcode `\\12\catcode `\$12\catcode
  `\&12\catcode `\#12\catcode `\^12\catcode `\_12\catcode `\%12\relax}%
\providecommand \@@startlink[1]{}%
\providecommand \@@endlink[0]{}%
\providecommand \url  [0]{\begingroup\@sanitize@url \@url }%
\providecommand \@url [1]{\endgroup\@href {#1}{\urlprefix }}%
\providecommand \urlprefix  [0]{URL }%
\providecommand \doibase [0]{http://dx.doi.org/}%
\providecommand \selectlanguage [0]{\@gobble}%
\providecommand \bibinfo  [0]{\@secondoftwo}%
\providecommand \bibfield  [0]{\@secondoftwo}%
\providecommand \BibitemOpen [0]{}%
\providecommand \EOS [0]{\spacefactor3000\relax}%
\providecommand \BibitemShut  [1]{\csname bibitem#1\endcsname}%
\let\auto@bib@innerbib\@empty

\newcommand{\cA}{{\cal A}}

\newcommand{\vphi}{\varphi}

\newcommand{\be}{\begin{equation}}      
\newcommand{\ee}{\end{equation}}      
\newcommand{\bea}{\begin{eqnarray}}      
\newcommand{\eea}{\end{eqnarray}}

\newcommand{\im}{\mathrm{i}}
\newcommand{\pf}{\mathop{\mathrm{Pf}}}

\begin{document}
\title{Kinetic derivation of generalized phase space Chern-Simons theory}

\preprint{RIKEN-QHP-250, RIKEN-STAMP-27}
\author{Tomoya Hayata}
\affiliation{
Department of Physics, Chuo University, 1-13-27 Kasuga, Bunkyo, Tokyo, 112-8551, Japan 
}
\author{Yoshimasa Hidaka}
\affiliation{
Theoretical Research Division, Nishina Center, RIKEN, Wako, Saitama 351-0198, Japan
}

\date{\today}

\begin{abstract}
We study anomalous transport phenomena induced by phase-space Berry curvature.
For that purpose we construct a kinetic theory in $2d$ phase space when all abelian Berry curvatures are nonzero.
We derive anomalous currents by calculating the complete form of the Poisson brackets of phase space coordinates.
Then we construct the low-energy effective theory to reproduce the anomalous currents obtained from the kinetic theory.
Such an effective theory is given by the Chern-Simons theory in $1+2d$ dimensions.
Some implications of the Chern-Simons theory are also discussed. 

\end{abstract}

\maketitle

\section{Introduction} 
Anomalous transport phenomena described by the Berry phase and the Berry curvature~\cite{Berry45} have attracted considerable attention in subdisciplines of physics.
For example, the Berry phase of electrons in Bloch states accounts for novel properties of insulators such as the adiabatic charge pumping in $1+1$ dimensions~\cite{PhysRevB.27.6083,PhysRevB.47.1651,RevModPhys.66.899}, 
and the quantum Hall effect in $1+2$ dimensions~\cite{PhysRevLett.45.494,PhysRevB.23.5632,PhysRevLett.49.405,PhysRevLett.51.51,PhysRevB.31.3372,RevModPhys.82.1959}. In the former case, electrons pick up the Berry phase calculated from a closed orbit on the torus defined by crystal momentum and time ($p,t$).
In the latter case, they pick up the one on the two-dimensional Brillouin zone ($p_x,p_y$).
Similarly, when parameters of a Hamiltonian depend on real-space coordinate $x$ due to inhomogeneous ordering~\cite{Fiebig,Kimura,Cheong} or strain spatial gradient~\cite{PhysRevB.34.5883}, the Bloch states depend on $x$, and then we can consider the Berry phase and the Berry curvature on the two-dimensional ($x,t$) or ($x,p$) space. 
In general, we can define the Berry curvature on any two-dimensional space projected from ($1+2d$)-dimensional phase space and time ($\bm x,\bm p,t$).
As shown below, the interplay of phase-space Berry curvatures becomes important as the dimension of the system increases.
Anomalous transport phenomena induced by such an interplay are potentially relevant in topological insulators~\cite{Bernevig1757,Konig766,Hsieh2008,Xia2009,Zhang2009,Chen2009,RevModPhys.82.3045}, or Dirac/Weyl semimetals~\cite{Liu864,PhysRevLett.113.027603,Xu613,Lu622,PhysRevX.5.031013} in 1+3 dimensions.
We can also study those in higher dimensions ($d\geq4$)~\cite{Zhang823,PhysRevLett.115.195303}, thanks to the invention of synthetic dimensions~\cite{PhysRevLett.108.133001}. 

There are two powerful approaches to describe the anomalous transport phenomena induced by the phase-space Berry curvature:
One is the kinetic theory, and the other is the topological field theory.
In the kinetic approach, transport phenomena are described by the classical dynamics of charge carriers.
In presence of the Berry curvature, their classical equation of motion is modified~\cite{PhysRevB.59.14915,RevModPhys.82.1959}, 
where additional Lorentz ``force"  for phase-space coordinates are induced by the Berry curvature. 
The effect of the Berry curvature on mixed space defined by real-space coordinate and momentum to anomalous transport phenomena has been studied by using this modified kinetic theory~\cite{Shindou2005399,PhysRevLett.102.087602}.
However, the Poisson brackets of phase space coordinates and the anomalous transport effects have been calculated only in the leading order of the derivative expansion when the mixed space Berry curvatures are nonzero. 
Namely, the full form of the Poisson brackets and the anomalous transport effects are not known when all ($1+2d$)-dimensional Berry curvatures are nonzero. 

The topological field theory gives the generating functional for electromagnetic responses of insulators, 
or more specifically of the system with no low-energy dynamical degrees of freedom in its bulk state to interact with external fields~\cite{PhysRevLett.62.82,PhysRevB.44.5246,ZHANG,PhysRevB.78.195424,RevModPhys.83.1057}.
We can discuss the topological field theory on phase space by considering the phase space as the manifold to define it. 
The arguments of topological field theory become electromagnetic gauge potentials and Berry connections~\cite{PhysRevB.78.195424,RevModPhys.83.1057}, namely, the connections in phase space. 
It has shown that this topological field theory on phase space is useful to describe anomalous transport phenomena induced by  the phase-space Berry curvature~\cite{PhysRevX.5.021018}.
However, the topological field theory on phase space was introduced through the analogy between the phase space theory and the noncommutative lowest Landau projected theory in $2d$ real space, and the direct connection between the original phase space theory, i.e., the aforementioned kinetic theory, and the topological field theory was unclear in Ref.~\cite{PhysRevX.5.021018}. 

In this paper, we study anomalous transport effects induced by the phase-space Berry curvature on the basis of the kinetic theory.
By elaborating the calculation of the Poisson brackets of phase space coordinates and the anomalous transport effects, we show that the kinetic theory is equivalent to the phase-space Chern-Simons theory given in Ref.~\cite{PhysRevX.5.021018}.
This paper is organized as follows:
In Sec.~\ref{sec:kinetic}, we construct a kinetic theory in the case that all abelian Berry curvatures in $1+2d$ dimensions are nonzero.
We derive the full form of the Poisson brackets of phase space coordinates, and calculate transport effects by using it.
In Sec.~\ref{sec:tft}, we construct a low energy effective theory to reproduce the transport phenomena obtained from the kinetic theory.
We show that the effective theory is, in fact, the generalized version of the phase-space Chern-Simons theory given in Ref.~\cite{PhysRevX.5.021018}.
We identify the energy $\varepsilon$ and momentum $p_i$ as gauge potentials ($A_t=-\varepsilon$ and $A_i=p_i$) 
as well as electromagnetic ones and Berry connections, and introduce the Chern-Simons theory on phase space with all those three gauge potentials. 
Some implications of the phase-space Chern-Simons theory including nonlinear electromagnetic responses are also discussed. 
Section~\ref{sec:concluding} is devoted to concluding remarks.

\section{Kinetic theory in the presence of Berry connections}
\label{sec:kinetic}
We consider the semiclassical dynamics of a Bloch electron, which has $2d$-dimensional phase space.
The action has the form~\cite{RevModPhys.82.1959}:
\be
{S} =\int dt \Bigl(\dot{\bm x}\cdot(\bm p+\bm A)+\dot{\bm p}\cdot\bm a-\varepsilon+A_t\Bigr),
\label{eq:action}
\ee
where $\varepsilon$ is an energy eigenvalue of a Bloch state $|u\rangle$, which can be labeled by time $t$ and phase space coordinates $\bm x=(x_1,\ldots,x_d)$ and $\bm p=(p_1,\ldots,p_d)$. 
$A_t$, $\bm A=(A_1,\ldots,A_d)$, and $\bm a=(a_1,\ldots,a_d)$ are the Berry connections and explicitly given as
$A_t=\langle u| \im\partial_t|u\rangle$, $\bm A=\langle u|\im\nabla_{\bm x}|u\rangle$, and $\bm a=\langle u|\im\nabla_{\bm p}|u\rangle$. 
The external electromagnetic fields can be absorbed into $A_t$ and $\bm A$.
In the following, we do not distinguish electromagnetic gauge potentials and Berry connections unless otherwise stated. 
We assume that all Berry connections fully depend on $t$, $\bm{x}$, and $\bm{p}$. We employ the Einstein convention for repeated indices. 
We introduce the (1+2d)-dimensional coordinates $\xi_\mu$ and the generalized connections $\cA_\mu$ as  $\xi_\mu= (\xi_0,\xi_a) = (t,\xi_a)$ and $\cA_\mu=(\cA_0,\cA_a)=(\cA_t,\cA_a)$ with $\xi_a=(\bm{x},\bm{p})$, $\cA_t=-\varepsilon+A_t$ and $\cA_a= (\bm{p}+\bm{A},\bm{a})$. 
We absorb the energy $\varepsilon$ and momentum $\bm p$ into gauge potentials~\cite{PhysRevX.5.021018}, 
which is different from the standard description of the kinetic theory~\cite{PhysRevB.59.14915,RevModPhys.82.1959}. 
Using these variables, we can write the action in the topological form:
\begin{equation}
\begin{split}
S = \int \cA_{\mu} d\xi_\mu.
\end{split}
\label{eq:action_topological}
\end{equation}
As will be seen below,  this topological nature implies that the low-energy effective field theory for transport phenomena 
is expressed as the Chern-Simons theory for $\cA_{\mu}$.

The equation of motion reads 
\be
\omega_{ab}\dot{\xi}_b= -\omega_{at}= \frac{\partial H}{\partial \xi_a}+\partial_t\cA_a,
\label{eq:ceom1}
\ee
where $H=-\cA_t=\varepsilon-A_t$, and we use $\dot\cA_a=\partial_t\cA_a+\dot{\xi}_b\partial_{\xi_b}\cA_a$. 
$\omega_{ab}$ and $\omega_{at}$ are the generalized Berry curvatures in phase space and time, and defined as 
\bea
\omega_{ab}&=&\frac{\partial \cA_b}{\partial \xi_a}-\frac{\partial \cA_a}{\partial \xi_b},
\label{eq:berry_curvature}\\
\omega_{at}&=&\frac{\partial \cA_t}{\partial \xi_a}-\frac{\partial \cA_a}{\partial t} .
\label{eq:berry_curvature_E}
\eea
We assume that $\det(\omega)$ is nonzero,
and then, the equation of motion~\eqref{eq:ceom1} is written as 
\be
\dot{\xi}_a=-\omega^{-1}_{ab}\omega_{bt} .
\label{eq:ceom2}
\ee
When we fix the gauge to satisfy $\partial_t\cA_a=0$, 
this can be expressed as the Hamilton equation,
\be
\dot{\xi}_a=\{\xi_a, H\}_p=\{\xi_a,\xi_b\}_p\frac{\partial H}{\partial \xi_b} ,
\label{eq:ceom3}
\ee
where $\{,\}_p$ represents the Poisson bracket.
From Eqs.~\eqref{eq:ceom2}, and~\eqref{eq:ceom3}, the Poisson bracket of phase-space coordinates reads
$\{\xi_a,\xi_b\}_p= \omega^{-1}_{ab}$.
At nonzero Berry curvatures, $\omega^{-1}_{ab}$, in general,  does not satisfy the standard canonical relations, so that $x_i$ and $p_i$ are no longer canonical pairs.
Also the invariant phase-space volume element is modified as $d^dxd^dp\sqrt{\det(\omega)}/(2\pi)^d$~\cite{RevModPhys.82.1959}.
Since $\omega$ is a skew symmetric matrix, its determinant is written as $\det(\omega)=\pf(\omega)^2$ by using the Pfaffian: 
\be
\pf(\omega)=\frac{1}{2^{d}d!}\epsilon_{a_1\cdots a_{2d}}\omega_{a_1a_2}\cdots\omega_{a_{2d-1}a_{2d}},
\ee
where $\epsilon_{a_1\cdots a_{2d}}$ is the totally anti-symmetric tensor ($\epsilon_{12\cdots2d}=1$). 
We note that the determinant of a real skew matrix is always nonnegative.
We can write the Pfaffian as 
\be
\pf(\omega)=\frac{1}{2^{d-1}(d-1)!}\epsilon_{aba_1\cdots a_{2d-2}}\omega_{ab}\omega_{a_1a_2}\cdots\omega_{a_{2d-3}a_{2d-2}} ,
\label{eq:pfaffian2}
\ee
where the summation over $a$ is not performed.
This corresponds to  the cofactor expansion of the Pfaffian.
We find that the inverse matrix is given as 
\be
\omega^{-1}_{ab}=\frac{1}{2^{d-1}(d-1)!\pf(\omega)}\epsilon_{baa_1\cdots a_{2d-2}}\omega_{a_1a_2}\cdots\omega_{a_{2d-3}a_{2d-2}} .
\label{eq:poisson_2d}
\ee
Let us check $\omega_{ca}\omega^{-1}_{ab}=\delta_{cb}$.
If $c=b$, from Eqs.~\eqref{eq:pfaffian2} and~\eqref{eq:poisson_2d}, $\omega_{ca}\omega^{-1}_{ab}$ equals unity.
If $c\neq b$, $\pf(\omega)\omega_{ca}\omega^{-1}_{ab}$ becomes the Pfaffian of another skew symmetric matrix $\tilde{\omega}$, which is obtained from $\omega$ by replacing the $b$-th row of $\omega$ by its $c$-th row.
The determinant of $\tilde{\omega}$ must be zero since two rows  ($b$- and $c$-th rows) are identical by its construction.
Since $\det(\tilde{\omega})=\pf(\tilde{\omega})^2=(\pf(\omega)\omega_{ca}\omega^{-1}_{ab})^2$, $\omega_{ca}\omega^{-1}_{ab}=0$ (we assume that $\det(\omega)\neq0$), and thus, $\omega_{ca}\omega^{-1}_{ab}=\delta_{cb}$.
Although the Poisson brackets of phase space coordinates are in general given by Eq.~\eqref{eq:poisson_2d},
it is useful for practical applications to show their explicit forms. We write down them in the case of $d=2,3$ in Appendices~\ref{sec:transports2}, and~\ref{sec:transports3}.

By using Eqs.~\eqref{eq:ceom2} and~\eqref{eq:poisson_2d}, we can discuss transport phenomena.
Under weak external fields, the transport phenomena can be described by the Boltzmann equation,
\be
\partial_t n(t,\xi_a)+\dot{\xi}_a\partial_{\xi_a}n(t,\xi_a)=0 ,
\label{eq:boltzmann_2d}
\ee
where $n(t,\xi_a)$ is the distribution function in $2d$-dimensional phase space, and we neglect collision terms.
$\dot{\xi}_a$ is given by the equation of motion~\eqref{eq:ceom2}, whose explicit form is
\be 
\sqrt{\det(\omega)}\dot{\xi}_a = \frac{(-1)^\nu}{2^{d-1}(d-1)!}\epsilon_{aba_1\cdots a_{2d-2}}\omega_{a_1a_2}\cdots\omega_{a_{2d-1}a_{2d-2}}\omega_{b t},
\label{eq:eom1_2d} 
\ee
where we introduce  the sign of the Pfaffian $(-1)^\nu= \pf(\omega)/\sqrt{\det(\omega)}$. In our kinetic regime, the sign is negative (see Appendices~\ref{sec:transports2}, and~\ref{sec:transports3} for the detail.)
The current can be calculated by averaging the velocity of quasiparticles, $\dot{\xi}_a$, over the phase space with the modified volume $\sqrt{\det(\omega)}/(2\pi)^d$. The current density is given as
\bea
j_a &=& \frac{1}{(2\pi)^d}\sqrt{\det(\omega)}\dot{\xi}_a n(t,\xi)
\notag \\
&=& \frac{(-1)^\nu}{(2\pi)^d2^{d-1}(d-1)!}\epsilon_{aba_1\cdots a_{2d-2}}\omega_{a_1a_2}\cdots\omega_{a_{2d-1}a_{2d-2}}\omega_{b t}n(t,\xi).
\label{eq:anomalouscurrent_2d}
\eea
Integrating Eq.~\eqref{eq:anomalouscurrent_2d} with respect to $p_i$, we obtain the current in real space.
Similarly, integrating Eq.~\eqref{eq:anomalouscurrent_2d} with respect to $x_i$, we obtain the current in momentum space.
Since the Boltzmann equation~\eqref{eq:boltzmann_2d} is symmetric in $x_i$ and $p_i$, we can consider the conservation law in momentum space, 
in which the force term $\sqrt{\omega}\dot{p}_in(t, \bm p)$ plays a role of the current. 
Equation~\eqref{eq:anomalouscurrent_2d} is valid not only at near equilibrium but also at far from equilibrium as long as the kinetic theory is applicable.
For example, we can calculate Berry curvature corrections to dissipative currents in nonequilibrium steady state with using the relaxation time approximation.

The local number density is given as $j_0=\sqrt{\det(\omega)}n(t,\xi)/(2\pi)^d$.
If we consider a band insulator ($n(t,\xi_a) =1$), we have 
\be
j_0(\xi )=\sqrt{\det(\omega)}/(2\pi)^d .
\label{eq:polarization_2d}
\ee
This is the local induced density, when the Berry curvatures are adiabatically introduced.
For the explicit form of Eqs.~\eqref{eq:anomalouscurrent_2d} and~\eqref{eq:polarization_2d} in $d=2,3$, see Appendices~\ref{sec:transports2}, and~\ref{sec:transports3}.

\section{Topological effective field theory}  
\label{sec:tft}
We here construct a low-energy effective theory to describe the reactions to Berry curvatures in Eqs.~\eqref{eq:anomalouscurrent_2d} and~\eqref{eq:polarization_2d} ($n(t,\xi_a) =1$).
We find that such an effective theory is represented by the phase space Chern-Simons theory~\cite{PhysRevX.5.021018}, 
whose action is
\be
S_{\rm CS}=\frac{(-1)^\nu}{(2\pi)^d (d+1)!}\int dtd^{2d}\xi\;\epsilon_{\mu_0\ldots \mu_{2d}}\cA_{\mu_0}\partial_{\mu_1}\cA_{\mu_2}\cdots  \partial_{\mu_{2d-1}}\cA_{\mu_{2d}} ,
\label{eq:CS_2d}
\ee
where $\mu_i=0,1,\ldots,2d$, $\partial_{\mu_i}=(\partial_t,\partial_{\xi_1}\ldots,\partial_{\xi_{2d}})$, $\cA_\mu=(\cA_t,\cA_a)=(-\varepsilon+A_t,p_i+A_i,a_i)$.
The current density is obtained by differentiating the action~\eqref{eq:CS_2d} with respect to $\cA_\mu$ as 
\be
\begin{split}
j_\mu &=\frac{\partial S_{\rm CS}}{\partial \cA_\mu}
\\
&=\frac{(-1)^\nu}{(2\pi)^d2^{d} d!}\epsilon_{\mu \mu_1\ldots \mu_{2d}}\omega_{\mu_1\mu_2}\cdots  \omega_{\mu_{2d-1}\mu_{2d}} ,
\end{split}
\label{eq:CS_current_2d}
\ee
where $\omega_{\mu_i\mu_j}=\partial_{\mu_i}\cA_{\mu_j}-\partial_{\mu_j}\cA_{\mu_i}$. 
Equation~\eqref{eq:CS_current_2d} recovers Eqs.~\eqref{eq:anomalouscurrent_2d} and~\eqref{eq:polarization_2d} ($n(t,\xi_a) =1$). 
This is our main result.
We see that all currents obtained from kinetic theory can be expressed by the Chern-Simons current~\eqref{eq:CS_current_2d}, so that two descriptions are equivalent.
We emphasize here that our kinetic theory can be applied to metals as well as insulators, and explains the anomalous transport effects in metals, which cannot be described by the phase-space Chern-Simons theory.
Below we discuss some interesting results obtained from Eq.~\eqref{eq:CS_current_2d}.
The following argument can be applied to both fermion and boson. 
This is because if we can prepare a uniformly filled band for bosons, the Boltzmann equation~\eqref{eq:boltzmann_2d} does not distinguish particle statistics without collision terms.
In fact such a band occupation of bosons is experimentally realized in ultracold atom systems~\cite{coldatom}. 

First we consider the adiabatic pumping current or equivalently the change of polarization in the presence of time-periodic perturbation.
We consider a single band insulator for simplicity. 
The distribution reads $n(t,\xi_a) =1$.
The change of polarization is obtained by integrating the current over the period of perturbation $T$ as, in $d=2$,
 \be
\begin{split}
P_i(x_i) &=-e\int_0^T dt j_i(x_i)
\\
&=e\int \frac{dtd^2p}{(2\pi)^2}\left[\delta_{ij}\Omega_{p_jt}+\left(\delta_{ij}\Omega_{p_kx_k}-\Omega_{p_{i}x_{j}}\right)\Omega_{p_jt}+\Omega_{p_ip_j}\Omega_{x_jt}\right],
\end{split}
\label{eq:polarization_2}
\ee
where $e>0$ is the electric charge, $\Omega_{p_jt}=\partial_{p_j} A_t- \partial_t a_j, \Omega_{p_ix_j}=\partial_{p_i} A_j- \partial_{x_i} a_j, \Omega_{p_ip_j}=\partial_{p_i}a_j-\partial_{p_j}a_i$, and $\Omega_{x_jt}=\partial_{x_j}A_t-\partial_t A_j.$
The first term is the adiabatic charge pumping given by Thouless~\cite{PhysRevB.27.6083}.
The second and third terms are corrections in the presence of spatial inhomogeneity~\cite{PhysRevLett.102.087602}. 
Equation~\eqref{eq:polarization_2} completely reproduces the result of Ref.~\cite{PhysRevLett.102.087602}. There exist no higher order corrections in $d=2$.
In contrast,  in $d=3$, we find
 \bea
P_i(x_i)=e\int \frac{dtd^3p}{(2\pi)^3}\Bigl[&&\left(\delta_{ij}+\delta_{ij}\Omega_{x_kp_k}-\Omega_{p_ix_j}-\Omega_{x_i}\Omega_{p_j}+\epsilon_{ikl}\epsilon_{j\bar{m}\bar{n}}\Omega_{p_{\bar{m}}x_k}\Omega_{p_{\bar{n}}x_l}/2\right)\Omega_{p_jt}
\notag \\
&&+ \left( \epsilon_{ijk}\Omega_{p_k}+\epsilon_{ijk}\Omega_{p_{\bar{k}}x_k}\Omega_{p_{\bar{k}}} \right)\Omega_{x_jt}
\Bigr] ,
\label{eq:polarization_3}
\eea
where $\Omega_{x_i}=\epsilon_{ijk}\Omega_{x_jx_k}/2$, and $\Omega_{p_i}=\epsilon_{ijk}\Omega_{p_jp_k}/2$.
This expression recovers the result of Ref.~\cite{PhysRevLett.102.087602} in the first order of the spatial gradient, 
and  the higher order terms are derived in Ref.~\cite{PhysRevX.5.021018}.

Since the Chern-Simons action is symmetric in $x_i$ and $p_i$, we can consider a momentum analogue of topological currents in real space.
As an example, let us consider the quantum Hall effect.  In $d=2$, the Chern-Simons action~\eqref{eq:CS_2d} contains the following term
\be
S_{\rm CS}=\frac{(-1)^\nu}{8\pi^2}\int dtd^2xd^2p\;\epsilon_{\mu_1\mu_2\mu_3}A_{\mu_1}\partial_{\mu_2} A_{\mu_3}\Omega_{p_1p_2} ,
\label{eq:CS_rp}
\ee
where $\mu_i=0,1,2$, $\partial_{\mu_i}=(\partial_t,\partial_{x_1},\partial_{x_2})$, and $A_\mu=(A_t,A_x,A_y)$.
$\Omega_{p_1p_2}=\partial_{p_1}a_{p_2}-\partial_{p_2}a_{p_1}$ is the conventional Berry curvature of a Bloch band.
If $A_i$ are gauge potentials for external electromagnetic fields, and $\Omega_{p_1p_2}$ does not depend on $x_i$, Eq.~\eqref{eq:CS_rp} 
reduces to
\be
S_{\rm CS}=\frac{C_1}{4\pi}\int dtd^2x\;\epsilon_{\mu_1\mu_2\mu_3}A_{\mu_1}\partial_{\mu_2} A_{\mu_3} ,
\label{eq:CS_r}
\ee
where
$C_1=(-1)^\nu\int d^2p\;\Omega_{p_1p_2}/2\pi$
is the first Chern number and takes only integer values.
This is the Chern-Simons theory for the integer quantum Hall effect. 
In addition, the Chern-Simons action~\eqref{eq:CS_2d} contains the following term
\be
S_{\rm CS}=\frac{(-1)^\nu}{8\pi^2}\int dtd^2xd^2p\;\Omega_{x_1x_2}\epsilon_{\mu_1\mu_2\mu_3}a_{\mu_1}\partial_{\mu_2} a_{\mu_3} ,
\label{eq:CS_pr}
\ee
where $\mu_i=0,1,2$, $\partial_{\mu_i}=(\partial_t,\partial_{p_1},\partial_{p_2})$, $a_\mu=(A_t,a_{p_1},a_{p_2})$, 
and $\Omega_{x_1x_2}=\partial_{x_1}A_{x_2}-\partial_{x_2}A_{x_1}$.
If $\Omega_{x_1x_2}$ does not depend on $p_i$, Eq.~\eqref{eq:CS_pr} reduces to
\be
S_{\rm CS}=\frac{C_p(-1)^\nu}{4\pi}\int dtd^2p\;\epsilon_{\mu_1\mu_2\mu_3}a_{\mu_1}\partial_{\mu_2} a_{\mu_3} ,
\label{eq:CS_p}
\ee
with 
$C_p=\int d^2 x\;\Omega_{x_1x_2}/2\pi$.
This is the magnetic flux perpendicular to the two dimensional system.
In the presence of topological defects with a magnetic charge such as a skyrmion, $C_p$ is quantized;
 the Chern-Simons action~\eqref{eq:CS_p} describes the integer quantum Hall effect in momentum space.
We can physically understand this effect through the adiabatic pumping.
Let us consider a time-periodic adiabatic perturbation. 
Since the current in momentum space is the acceleration, the time and momentum integration of the quantum Hall current, $j_{p_i}=(C_p/2\pi)\epsilon_{ij}\Omega_{p_jt}$, 
over a period of cycle and Brillouin zone gives a contribution to the change of total momentum over one cycle. 
In fact, there is a leading contribution, which is momentum analogue of the Thouless pumping~\cite{PhysRevLett.109.010601}.
Equation~\eqref{eq:CS_p} gives the correction to the adiabatic momentum pumping like the last term in Eq.~\eqref{eq:polarization_2}.
The full form of the momentum pumping is given by Eqs.~\eqref{eq:polarization_2} and~\eqref{eq:polarization_3} with the exchange of $x$ and $p$. 

\section{Concluding remarks} 
\label{sec:concluding}
We have studied anomalous transport effects induced by the Berry curvature on the basis of the kinetic theory and the phase space Chern-Simons theory.
To discuss the  transport phenomena by using the kinetic theory, we have derived the exact form of Poisson brackets of phase space coordinates.
By using it, we have analyzed anomalous transport phenomena.
Then, we have shown that all transport phenomena calculated from the kinetic theory can be described by the generalized version of the phase space Chern-Simons theory.
We  have shown a clear connection between the kinetic theory and the phase space Chern-Simons theory, which was not explicitly derived in the pervious work~\cite{PhysRevX.5.021018}.
Our work also provides the clear derivation of the generalized phase space Chern-Simons theory, not based on the analogy between the phase space theory and the noncommutative lowest Landau projected theory in $2d$ real space~\cite{PhysRevX.5.021018}.
The explicit form of anomalous transport effects, which are important for practical purpose, are summarized in Appendices~\ref{sec:transports2}, and~\ref{sec:transports3}. 

We comment on the fact that the Chern-Simons current~\eqref{eq:CS_current_2d} includes nonlinear responses to Berry curvatures, 
which are higher $\hbar$ order terms, and in general, suffer from the next leading order corrections in the view of kinetic theory~\cite{PhysRevLett.102.087602}. 
Such corrections will disappear if the kinetic theory with higher $\hbar$ order terms still keeps the topological form like Eq.~\eqref{eq:action_topological}. Namely, the low-energy theory needs to have area-preserving diffeomorphism invariance in phase space.  
In the presence of the invariance, corrections are restricted to the form described by the phase space Chern-Simons theory. 
However, the condition for the system to have phase-space area-preserving diffeomorphism invariance as a low-energy emergent symmetry
has not been discussed so far, and to clarify it is an important future study.

There are several generalizations of our work.
One direction is to consider degenerate Bloch states and non-abelian Berry curvatures.
In this case, a non-abelian version of the generalized phase-space Chern-Simons theory 
would give us the inclusive description of electromagnetic properties, 
which should include the unified theory of topological insulators given in Ref.~\cite{PhysRevB.78.195424}.
Another direction is a generalization to include other transport phenomena such as thermal or spin currents~\cite{PhysRevLett.69.953,Wen_thermal,PhysRevLett.114.016805,PhysRevLett.95.057205}.
It is also interesting to consider the dynamical gauge fields in phase space. 
We assume that all gauge potentials in phase space are static in this paper, but it could be dynamical if it arises from the corrective behavior of interacting many-body electrons. 
In fact, such a dynamical gauge fields would emerge 
when we consider an low-energy effective theory of the fractional quantum Hall state~\cite{PhysRevLett.62.82,PhysRevB.44.5246,ZHANG}.
It will be interesting to see the behavior of the effective theory obtained by integrating out the dynamical gauge fields from the generalized phase space Chern-Simons theory as conducted in the fractional quantum Hall effect.

\begin{acknowledgements}

This work was supported by JSPS Grant-in-Aid for Scientific Research (No: JP16J02240).
This work was also partially supported by JSPS KAKENHI Grants Numbers 15H03652, 16K17716 and the RIKEN interdisciplinary Theoretical Science (iTHES) project.
\end{acknowledgements}

\appendix
\section{Phase-space Berry curvature}
\label{sec:berry}
Here we consider a Weyl Hamiltonian in $d=3$ as an example to have nonzero abelian phase-space Berry curvatures.
We start with the general $2\times2$ Weyl Hamiltonian
\be
{\cal H} = R_x\sigma_x+R_y\sigma_y+R_z\sigma_z,
\label{eq:Weyl_Hamilatonian}
\ee
where $R_i$ are the function of spatial coordinates $\bm x$, momentum $\bm p$, and time $t$.
$\sigma_i$ are the Pauli matrices.
The instantaneous eigenvalue of Eq.~\eqref{eq:Weyl_Hamilatonian} is given as 
$\varepsilon_{\pm}=\pm\sqrt{R_x^2+R_y^2+R_z^2}$.
The eigenvectors read
\be
u^{-}(p)=  
\begin{pmatrix}
\sin\frac{\theta}{2}e^{-i\vphi} \\
-\cos\frac{\theta}{2}
\end{pmatrix},\;
u^{+}(p)=  
\begin{pmatrix}
\cos\frac{\theta}{2}e^{-i\vphi} \\
\sin\frac{\theta}{2}
\end{pmatrix},
\label{eq:Rashba_wave}
\ee
where we used the spherical coordinates in $\bm R$ space, namely, $\theta$ and $\varphi$ are defined as
\be
\tan \theta=\frac{\sqrt{R_x^2+R_y^2}}{R_z} ,\;\;
\tan \vphi= \frac{R_y}{R_x} .
\ee
We define the Berry connections and the Berry curvatures in $\bm R$ space as 
\be
 \bm a_R^{\pm}(\bm R)  =   \left(u^{\pm}(\bm R)\right)^\dagger\im\nabla_{R}u^{\pm}(\bm R) .
\ee
We find that the Berry curvature in $\bm R$ space has a monopole at $\bm R=0$:
\be
\bm\Omega_R^\pm = \nabla_{R}\times \bm a_R^{\pm}
= \mp\frac{1}{2}\frac{\bm R}{|\bm R|^3}.
\label{eq:Berry_R}
\ee
Now the Berry curvatures in phase space can be calculated by using the so called pull back from $\bm R$-space as
\bea
\Omega_{p_ip_j}^{\pm} 
&=& \frac{1}{2}\partial_{p_i} R^m\partial_{p_j} R^l\epsilon^{mln}\Omega_R^{\pm n} ,
\label{eq:Berry_curvature1}
\\
\Omega_{x_ix_j}^{\pm} 
&=& \frac{1}{2}\partial_{x_i} R^m\partial_{x_j} R^l\epsilon^{mln}\Omega_R^{\pm n} ,
\label{eq:Berry_curvature2}
\\
\Omega_{x_ip_j}^{\pm} 
&=& \frac{1}{2}\partial_{x_i} R^m\partial_{p_j} R^l\epsilon^{mln}\Omega_R^{\pm n} ,
\label{eq:Berry_curvature3}
\\
\Omega_{x_it}^{\pm} 
&=& \frac{1}{2}\partial_{x_i} R^m\partial_{t} R^l\epsilon^{mln}\Omega_R^{\pm n} ,
\label{eq:Berry_curvature4}
\\
\Omega_{p_it}^{\pm} 
&=& \frac{1}{2}\partial_{p_i} R^m\partial_{t} R^l\epsilon^{mln}\Omega_R^{\pm n} .
\label{eq:Berry_curvature5}
\eea
Since $\bm R$ usually depends on momentum, i.e., $\nabla_p \bm R\neq0$, if $\nabla_x \bm R\neq0$, we have nonzero mixed space Berry curvature $\Omega_{x_ip_j}^{\pm}$. 
For example in the linear Weyl model with $R_i=v_i(p_i-p_{wi})$, where $v_i$, and $p_{wi}$ are the Fermi velocity and the position of the Weyl node,
 $\nabla_x \bm R\neq0$ can be achieved by making $v_i$ or $p_{wi}$ spatially inhomogeneous.
 In terms of physical process, this is realized e.g., by applying strain gradient.

\section{Poisson brackets in $1+2$ dimensions}
\label{sec:transports2}
Here we derive the Poisson brackets in $d=2$ by assuming that all abelian Berry curvatures become nonzero.
We use the standard notation to compare the results with previous works. 
For the phase space coordinates $\xi_a=(x_i,p_i)=(x_1,x_2,p_1,p_2)$, the equation of motion reads 
\be
\left(\Omega-J\right)_{ab}\dot{\xi}_b=\frac{\partial H}{\partial \xi_a}+\partial_t \cA_a,
\label{eq:ceom1_standard}
\ee
where $H=\varepsilon-A_t$, and $J_{ab}$ is an anti-symmetric matrix:
\be
J_{ab}=
\begin{pmatrix}
0 & I \\
-I & 0 
\end{pmatrix} ,
\ee
with $I$ being the $2\times 2$ unit matrix.
$\Omega_{ab}$ is the Berry curvature  and defined as 
\be
\Omega_{ab}=\frac{\partial \cA_b}{\partial \xi_a}-\frac{\partial \cA_a}{\partial \xi_b} ,
\label{eq:berry_curvature_standard}
\ee
with $\cA_a=(\bm A_i,\bm a_i)$.
We assume $\det(\Omega-J)\neq0$. Then the equation of motion reads 
\be
\dot{\xi}_a=\left(\Omega-J\right)^{-1}_{ab}\left(\frac{\partial H}{\partial \xi_b} +\partial_t\cA_b\right).
\label{eq:ceom2_standard}
\ee
From this equation, the Poisson brackets of phase space coordinates are
\be
\{\xi_a,\xi_b\}_p= \left(\Omega-J\right)^{-1}_{ab}.
\label{eq:poisson_standard}
\ee
The invariant phase space volume element is modified as $d^2xd^2p\sqrt{\det(\Omega-J)}/(2\pi)^2$.
In $1+2$ dimensions, $\Omega-J$ is given explicitly as
\be
\left(\Omega-J\right)_{ab}=
\begin{pmatrix}
 \Omega_{x_ix_j} & -\delta_{ij}-\Omega_{p_jx_i} \\
 \delta_{ij}+\Omega_{p_ix_j} & \Omega_{p_ip_j} \\
 \end{pmatrix} ,
\ee
where $i,j=1,2$.
Its inverse matrix is 
\be
\left(\Omega-J\right)^{-1}_{ab}=
\frac{(-1)^\nu}{\sqrt{\det(\Omega-J)}}
\begin{pmatrix}
- \Omega_{p_ip_j} & -\delta_{ij}(1+\Omega_{p_kx_k})+\Omega_{p_{i}x_{j}} \\
 \delta_{ij}(1+\Omega_{p_kx_k})-\Omega_{p_jx_i} & -\Omega_{x_ix_j} \\
 \end{pmatrix}.
\label{eq:inverse_2d}
\ee
In the kinetic regime $\Omega\ll 1$, the sign of the Pfaffian is negative: $(-1)^\nu=\pf(\Omega-J)/\sqrt{\det(\Omega-J)}\simeq \pf(-J)=-1$,
and the Jacobian of $\Omega-J$ reads
\be
\sqrt{\det (\Omega-J)}=1+\Omega_{p_ix_i}-\epsilon_{abcd}\Omega_{\xi_a\xi_b}\Omega_{\xi_c\xi_d}/8 ,
\ee 
where $\epsilon_{abcd}$ is the totally anti-symmetric tensor with $\epsilon_{1234}=1$ ($\xi_a=(\xi_1,\xi_2,\xi_3,\xi_4)$).

From Eq.~\eqref{eq:inverse_2d}, the Poisson brackets are  expressed as
\bea
\{x_i,x_j\}_p &=& \Omega_{p_ip_j}/\sqrt{\det(\Omega-J)},
\\
\label{eq:poisson2_xx}
\{x_i,p_j\}_p &=& \left(\delta_{ij}(1+\Omega_{p_kx_k})-\Omega_{p_ix_j}\right)/\sqrt{\det(\Omega-J)},
\\
\label{eq:poisson2_xp}
\{p_i,x_j\}_p &=& -\left(\delta_{ij}(1+\Omega_{p_kx_k})-\Omega_{p_jx_i}\right)/\sqrt{\det(\Omega-J)},
\\
\label{eq:poisson2_px}
\{p_i,p_j\}_p &=& \Omega_{x_ix_j}/\sqrt{\det(\Omega-J)}.
\label{eq:poisson2_pp}
\eea
From Eqs.~\eqref{eq:ceom2_standard} and~\eqref{eq:inverse_2d}, we have
\bea
\sqrt{\det(\Omega-J)}\dot{x}_i &=& \left(\delta_{ij}(1+\Omega_{p_kx_k})-\Omega_{p_ix_j}\right)\tilde{v}_j
-\Omega_{p_ip_j}\widetilde E_j,
\label{eq:eom1_2d} \\
\sqrt{\det(\Omega-J)}\dot{p}_i &=& 
 \left(\delta_{ij}(1+\Omega_{p_kx_k})-\Omega_{p_jx_i}\right) \widetilde E_j
+\Omega_{x_ix_j}\tilde{v}_j ,
\label{eq:eom2_2d}
\eea
where $\tilde{v}_i=\partial_{p_i} \varepsilon-\Omega_{p_it}$, and $\widetilde{E}_i=\Omega_{x_it}-\partial_{x_i} \varepsilon$~\cite{RevModPhys.82.1959}.
At finite magnetic fields perpendicular to the two-dimensional system ($\Omega_{x_1x_2}\neq0$), 
the energy of a quasi-particle has a correction due to its nonzero magnetic moment, 
which is given as $\varepsilon =\varepsilon_0\left(1- \frac{q}{\hbar c}\Omega_{x_1x_2}\Omega_{p_1p_2}\right)$~\cite{RevModPhys.82.1959} with $\varepsilon_0$ being the energy at zero magnetic field.
When the magnetic field is spatially non-uniform, it induces an effective electric field through $\partial_{x_i} \varepsilon$.
For a band insulator, $n(t,\xi_a)=1$,  we can calculate the anomalous current  in real space by integrating Eq.~\eqref{eq:eom1_2d} over momentum space as
 \be
\begin{split}
j_i
&=\int \frac{d^2p}{(2\pi)^2}\sqrt{\det(\Omega-J)}\dot{x}_i
\\
&=\int \frac{d^2p}{(2\pi)^2}\left[\left(\delta_{ij}(1+\Omega_{p_kx_k})-\Omega_{p_{i}x_{j}}\right)\tilde{v}_j-\Omega_{p_ip_j}\widetilde E_j \right] .
\end{split}
\ee

\section{Poisson brackets in $1+3$ dimensions}
\label{sec:transports3}
Here we derive the Poisson brackets of phase space coordinates in $1+3$ dimensions by assuming that all abelian Berry curvatures become nonzero.
In $1+3$ dimensions, $\Omega-J$ reads
\be
\left(\Omega-J\right)_{ab}=
\begin{pmatrix}
 \Omega_{x_ix_j} & -\delta_{ij}-\Omega_{p_j x_i} \\
 \delta_{ij}+\Omega_{p_ix_j} & \Omega_{p_ip_j} \\
 \end{pmatrix} ,
\ee
where $i,j=1,2,3$. 
Its inverse matrix is given as (see also the Poisson brackets shown below)
\be
\begin{split}
&\sqrt{\det (\Omega-J)}\left(\Omega-J\right)^{-1}_{ab}= \\
&
{\scriptsize
\begin{pmatrix}
 \epsilon_{ijk}\Omega_{p_k}+\epsilon_{ijk}\Omega_{p_{\bar{k}}x_k}\Omega_{p_{\bar{k}}} 
 &  +\delta_{ij}(1+\Omega_{p_kx_k})-\Omega_{p_i x_j}-\Omega_{x_i}\Omega_{p_j}+\epsilon_{ikl}\epsilon_{j\bar{m}\bar{n}}\Omega_{p_{\bar{m}}x_k}\Omega_{p_{\bar{n}}x_l}/2 \\
  -\delta_{ij}(1+\Omega_{p_kx_k})+\Omega_{p_jx_i}+\Omega_{x_j}\Omega_{p_i}
  -\epsilon_{jkl}\epsilon_{i\bar{m}\bar{n}}\Omega_{p_{\bar{m}}x_k}\Omega_{p_{\bar{n}}x_l}/2 
 & \epsilon_{ijk}\Omega_{x_k}+\epsilon_{ij\bar{k}}\Omega_{p_{\bar{k}}x_l}\Omega_{x_l} \\
 \end{pmatrix} ,
}
\end{split}
\ee
where $\Omega_{x_i}=\epsilon_{ijk}\Omega_{x_jx_k}/2$, and $\Omega_{p_i}=\epsilon_{ijk}\Omega_{p_jp_k}/2$. $\epsilon_{ijk}$ is the totally anti-symmetric tensor with $\epsilon_{123}=1$. 
We used the fact that $(-1)^\nu=-1$.
The Jacobian of $\Omega-J$ reads
\bea
\sqrt{\det (\Omega-J)}&=& 1+\Omega_{p_ix_i}-\Omega_{x_i}\Omega_{p_i}+\left((\Omega_{p_ix_i})^2-\Omega_{p_ix_j}\Omega_{p_jx_i}\right)/2 
\notag \\
&&-\Omega_{p_i}\Omega_{p_i x_j}\Omega_{x_j}+\epsilon_{ijk}\epsilon_{\bar{l}\bar{m}\bar{n}}\Omega_{p_{\bar{l}}x_i}\Omega_{p_{\bar{m}}x_j}\Omega_{p_{\bar{n}}x_k}/6 .
 \eea
The detail of the Jacobian is not important to calculate transport phenomena since we only need numerators of the Poisson brackets of phase space coordinates;
they are given as
\bea
\{x_i,x_j\}_p &=& \left( \epsilon_{ijk}\Omega_{p_k}+\epsilon_{ijk}\Omega_{p_{\bar{k}}x_k}\Omega_{p_{\bar{k}}} \right)/\sqrt{\det (\Omega-J)},
\label{eq:poisson3d_xx}
\\
\{x_i,p_j\}_p &=& \left(\delta_{ij}(1+\Omega_{p_kx_k})-\Omega_{p_ix_j}-\Omega_{x_i}\Omega_{p_j}+\epsilon_{ikl}\epsilon_{j\bar{m}\bar{n}}\Omega_{p_{\bar{m}}x_k}\Omega_{p_{\bar{n}}x_l}/2\right)/\sqrt{\det (\Omega-J)},
\label{eq:poisson3d_xp}
\\
\{p_i,x_j\}_p &=&-\left(\delta_{ij}(1+\Omega_{p_kx_k})-\Omega_{p_jx_i}-\Omega_{x_j}\Omega_{p_i}+\epsilon_{jkl}\epsilon_{i\bar{m}\bar{n}}\Omega_{p_{\bar{m}}x_k}\Omega_{p_{\bar{n}}x_l}/2\right)/\sqrt{\det (\Omega-J)},\;\;\;\;\;\;
\label{eq:poisson3d_px}
\\
\{p_i,p_j\}_p &=& \left(\epsilon_{ijk}\Omega_{x_k}+\epsilon_{ij\bar{k}}\Omega_{p_{\bar{k}}x_l}\Omega_{x_l}\right)/\sqrt{\det (\Omega-J)}.
\label{eq:poisson3d_pp}
\eea
When $\Omega_{x_ip_j}$ are set to zero, these results recover the expressions given in Ref.~\cite{RevModPhys.82.1959}. 
From the Poisson brackets~\eqref{eq:poisson3d_xx}-\eqref{eq:poisson3d_pp},
we have 
\bea
\sqrt{\det (\Omega-J)}\dot{x}_i &=& \left(\delta_{ij}(1+\Omega_{p_kx_k})-\Omega_{p_ix_j}-\Omega_{x_i}\Omega_{p_j}+\epsilon_{ikl}\epsilon_{j\bar{m}\bar{n}}\Omega_{p_{\bar{m}}x_k}\Omega_{p_{\bar{n}}x_l}/2\right)\tilde{v}_j
\notag \\
&&- \left( \epsilon_{ijk}\Omega_{p_k}+\epsilon_{ijk}\Omega_{p_{\bar{k}}x_k}\Omega_{p_{\bar{k}}} \right)\widetilde{E}_j
\label{eq:eom3d1} \\
\sqrt{\det (\Omega-J)}\dot{p}_i &=& 
\left(\delta_{ij}(1+\Omega_{p_kx_k})-\Omega_{p_jx_i}-\Omega_{x_j}\Omega_{p_i}+\epsilon_{jkl}\epsilon_{i\bar{m}\bar{n}}\Omega_{p_{\bar{m}}x_k}\Omega_{p_{\bar{n}}x_l}/2\right)
\widetilde {E}_j
\notag \\
&&+\left(\epsilon_{ijk}\Omega_{x_k}+\epsilon_{ij\bar{k}}\Omega_{p_{\bar{k}}x_l}\Omega_{x_l}\right)\tilde{v}_j
\label{eq:eom3d2}
\eea
where $\tilde{v}_i=\partial_{p_i} \varepsilon-\Omega_{p_it}$, and $\widetilde{E}_i=\Omega_{x_it}-\partial_{x_i} \varepsilon$~\cite{RevModPhys.82.1959}.
We can calculate the anomalous current in real space from $j_i(t,\bm x)=\int d^3p/(2\pi)^3\sqrt{\det (\Omega-J)}\dot{x}_i n(t, \bm{x}, \bm{p})$.
Also the change of polarization $P_i$ is calculated from $ P_i=\int dt\int d^3x  j_i$, which is discussed in the main text.


\bibliography{./cs}

\end{document}